\documentstyle[prl,amssymb,aps,epsfig,epsf,multicol]{revtex}

\begin{document}

\draft

\title{Fluctuations and Market Friction in Financial Trading}

\author{Bernd Rosenow}

\address{Institut f\"ur Theoretische Physik, Universit\"at zu K\"oln,
D-50937 K\"oln, Germany  }

\date{\today}

\maketitle

\begin{abstract}
  We study the relation between stock price changes and the 
difference in the number of sell and buy orders. Using a soft spin 
  model, we describe the price impact of order imbalances and find an
  analogy to the fluctuation-dissipation theorem in physical systems.
  We empirically investigate  fluctuations and market friction for a
  major US stock and find support for our model calculations.
\end{abstract}



\begin{multicols}{2}
  
  The unpredictable up and down movements in the stock market have
  always captured the interest and imagination of investors.  The
  scientific investigation of these phenomena started with Bachelier's
  comparison of stock price dynamics with a random walk \cite{Bach00}.
  This study has been refined in many respects \cite{CaLoMa97}.
  Interest has been devoted to the precise shape of the distribution
  of returns (difference of the logarithm of stock prices at different
  times), which is characterized by a high probability for large
  fluctuations \cite{Mandel63,StMa95,Lux96,GoMeAmSt98}.
  
  After it was realized that large price fluctuations tend to cluster
  together in time, stock price returns were described by models of
  volatility (standard deviation of returns) changing in time
  \cite{Schwert89,volreview}. This effect is captured in time series
  models, in which the volatility at a given time depends on the
  magnitude of previous returns \cite{Engle82,Bollerslev86}.  It has
  been actively investigated how the stochastic properties of price
  dynamics can be related to the market microstructure, i.e.  the
  rules and motivations according to which agents act in a financial
  market. Although the details of models differ
  \cite{LeLeSo,ChZh,CoBo97,LuMa99,Farmer98} , they have been
  successful in reproducing the empirical observations.

Here, we follow a different approach motivated by the successful
application of physics concepts to study the economy
\cite{Farmer99,BoPo00,MaSt00}. We investigate the price dynamics on an
intermediate or coarse grained time scale $\Delta t$, which is long
enough to average over the details of market microstructure, but short
enough to resolve the trading dynamics. In physical systems,
there exists a quantitative relation between the strength of
fluctuations of an observable  and the rate, at which energy is dissipated
when the same observable is pulled by an external force. Einstein
discovered this relation first, when he studied the fluctuations of
small particles in a fluid \cite{Einstein05}. In this case, the
variance of position fluctuations is proportional to temperature times
the mobility of particles. This relation was later on generalized to a
wide class of systems \cite{CaWe52} and is known as fluctuation-dissipation
theorem. 

In this letter, we use an effective model for stock price dynamics to
derive a similar relation between the time varying volatility of stock
returns and the ``friction constant for stock prices'' or market
liquidity. Market liquidity relates returns to the difference between
buy and sell orders (order imbalance), which acts as an external
force. We study the dependence of stock price changes on order
imbalance empirically by using the method of data analysis and some of the 
results of \cite{private,response}. We find that the empirical results
agree well with our model.

{\it Model calculation:} The observable quantity we are interested in
is the  logarithmic stock price changes within a time interval $\Delta t$

\begin{equation}
G_{\Delta t}(t) = \ln S(t) -\ln S(t-\Delta t) \ ,
\end{equation}
  
where $S(t)$ is the price of a given stock at time $t$.  Transaction
prices at a stock exchange lie usually in a finite interval between
the bid price (the price a trader offers to pay for a 
stock) and the ask price (the price at which a dealer is willing to
buy the stock). In addition, the historical prices studied take only
discrete (tick) values. This motivates to model price changes by a
spin model, and for the virtue of easier analytical calculations by a
soft spin model \cite{Ma76}.
  
The spin variable we use is the ``instantaneous return'' $g(t) = \tau
{d\over dt} \ln S(t)$, where the average time interval between trades
$\tau$ sets the time scale of the problem. The observable returns on time
scale $\Delta t$ are related to them via
%
\begin{equation}
G(t)_{\Delta t} = {1 \over \tau} \int_{t - \Delta t}^t dt^\prime g(t^\prime)
\ .
\end{equation}
%
In an analogous way, we define an instantaneous order imbalance $q(t)=
\tau {d\over d t} Q(t)$.  Here, order imbalance is the number of buy
orders minus the number of sell orders divided by the total number of
outstanding shares of a given stock.  We describe the dynamics of
$g(t)$ by the following stochastic differential equation
%
\begin{eqnarray}
\tau \partial_t g(t)= - r  g(t) - \kappa g^3(t) + \mu q(t)     
+ \xi_i(t)  \ \ \ .
\label{Langevin}
\end{eqnarray}
%
Mathematically, the $r$- and $\kappa$- term enforce the soft spin
constraint of finite and discrete price changes. From a finance point
of view, the linear term describes relaxation and controls the strength of
fluctuations.  The cubic term stabilizes the theory in a regime of
strong fluctuations (small $r$). The order imbalance $q(t)$ is
analogous to the magnetic field in the theory of spins, its coupling
constant $\mu$ is in analogy to the magneton in magnetic systems.  The
Gaussian white noise $\xi(t)$ has the correlator $\langle \xi(t)
\xi(t^\prime)\rangle = \tau \delta(t-t^\prime)$. It represents the
influx of news not captured by the news content of the order
imbalance. Stochastic differential equations for stock
price changes with a quadratic instead of the cubic term were studied 
in \cite{Farmer98,CoBo99}. A generalization of Eq.(\ref{Langevin}) to
many stocks was used to explain correlations between them \cite{sectors}.

The variance of price changes on a time scale $\Delta t \gg \tau$ is well
approximated by 
%
\begin{eqnarray}
\langle (G_i(\Delta t))^2 \rangle_0 =  
{2 \over r^2 }\ {1 \over \tau} \ 
\Delta t \ \ ,
\label{vol.eq}
\end{eqnarray}
%
where the average is taken without the cubic term. This approximation
is valid as long as the system is not in the vicinity of a critical
point.  Eq.(\ref{vol.eq}) describes a diffusion process with diffusion
constant $D={2 \over r^2 \tau}$. In order to describe the stochastic
volatility and fat tails observed in empirical data, we must allow for
a time dependence of both the frequency of trades $1 \over \tau$ and
the standard deviation of the price of individual transactions.  In
Ref. \cite{diffusion} it was shown that price changes of financial
transactions can be consistently interpreted in such a framework.
When $r$ becomes a random variable, is causes multiplicative noise in
Eq.(\ref{Langevin}), which can account for the power law tails of the
return distribution \cite{multnoise,TaTa99}.

\begin{figure}
\vspace*{.5cm}
\hspace*{-1cm}
\epsfig{file=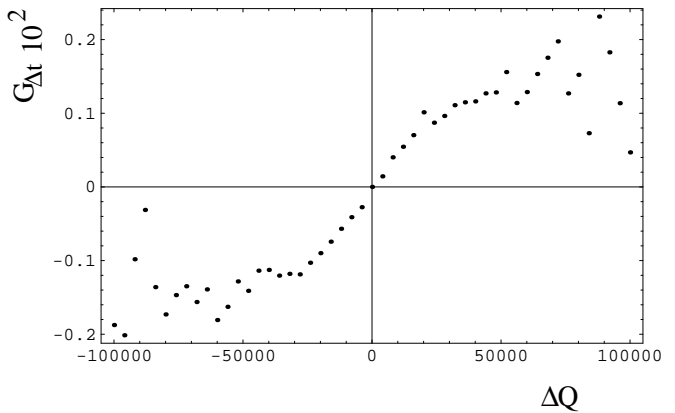,width=6cm,angle=270}
\caption{Price impact function: expectation value of
  return $\langle G_{\Delta t}\rangle_{\Delta Q}$ for an order
  imbalance $\Delta Q$ plotted against that order imbalance $[21,22]$.
  Returns are measured in percent, the order
 imbalance in number of stocks.  Note the linear part for small
  $\Delta Q$.  }
\label{chi.f}
\end{figure}

The price impact function describes the functional relation between
the expectation value $\langle G_{\Delta t}\rangle_{Q_{\Delta t}}$
conditioned on the volume imbalance and the volume imbalance itself.
We define the susceptibility $\chi$ as the slope of the price impact function
close to zero volume imbalance
%
\begin{eqnarray}
\chi = \lim_{\Delta Q \to 0}
{\langle  G_{\Delta t}\rangle_{\Delta Q} \over \Delta Q} \ .
\end{eqnarray}
%
As stocks are traded in discrete units, this limit can in an
empirical study only be approximated. The inverse of this
susceptibility can be interpreted as market liquidity.  For an order
imbalance which is constant in the time interval $\Delta t$ the full
price impact function can be calculated analytically for the model
Eq.(\ref{Langevin}) and compares well  to the empirical data in Fig.
\ref{chi.f}. Such a situation could be realized if an informed trader
breaks up a large trade into many small ones and trades at a fixed
frequency.  

From Eq.(\ref{Langevin}) we find $\chi = \mu/r$. Combining
this with Eq.(\ref{vol.eq}) for the volatility, we obtain for the
linear regime of small fluctuations
%
\begin{eqnarray}
\sqrt{\langle (G_{\Delta t})^2\rangle } =
\sqrt{2  \Delta t / \tau}\  \ \chi/\mu\ .
\label{flucdiss.eq}
\end{eqnarray}
%
This relation and its empirical test are our central results.  The
square root on the r.h.s. is essentially the square root of the number
of trades in the time interval $\Delta t$ and in analogy to the
temperature in physical systems.  The relation Eq.(\ref{flucdiss.eq})
differs from the Einstein relation in that the standard deviation of
fluctuations appears on the l.h.s.  instead of the variance. The
reason for this is the difference in ``experimentally'' relevant time
scales: as $\Delta t$ was assumed to be much larger than the
microscopic time scale $\tau$ in our coarse grained description of
stock price dynamics, $G_{\Delta t}(t)$ is proportional to the zero
frequency Fourier component of the instantaneous return, and not to
the integral over all frequency components as in the derivation of the
Einstein relation.

{\it Empirical Results:} The empirical test of Eq.(\ref{flucdiss.eq})
uses the method of data analysis and some of the results of
\cite{private,response}. The present author got to know about the shape
of the empirical price impact function and its similarity to the 
magnetization curve of spin systems from the authors of 
\cite{private,response}, and only later on learned about related work in
the economics literature \cite{KeKo}.-

In the framework of the efficient market
hypothesis the price of a stock reflects present and discounted future
earnings of the underlying company. The estimation of these earnings
depends on information about that company. Information may be
disseminated by a news release, a press article, or by rumors.
In addition, traders extract information about a company (or at least
information about the beliefs of other traders concerning that
company) from the order imbalance. This is  reasonable under the assumption
that somebody selling the stock might do that because he has superior
knowledge about the underlying company and expects its value to drop.
Hence, one can assume that the order imbalance will have impact on the
price of a stock.

We analyze stock prices of General Electric, a major American company
for the year 1997. Price, volume, bid and ask price of all
transactions of this stock are recorded in the Trades and Quotes (TAQ)
data base published by the New York Stock exchange. We analyzed a
total number of $10^6$ transactions contained in that data base.  The
question, whether a trade is buyer or seller initiated is difficult to
answer, as there is a buyer and a seller for each trade. We try to
answer it by finding out which one is more eager to do the trade, i.e.
we compare the transaction price to the bid and ask price
\cite{private,response,KeKo,HaLoMa92,BlMaTe}.  First, we try to match
the price of a trade with quotes at the same stock exchange with a
time stamp at least two seconds prior to the transaction time. In this
way we account for possible time lags in the reporting of prices. If
the trade occurs at the bid price, it is classified as seller
initiated, if it occurs between bid and ask price it is classified as
neutral, and a trade at the ask price is classified as a buyer
initiated. If the transaction price is lower than the bid price or
higher than the ask price the trade is discarded as not matched.

\begin{figure}
\vspace*{.5cm}
\hspace*{-1cm}
\epsfig{file=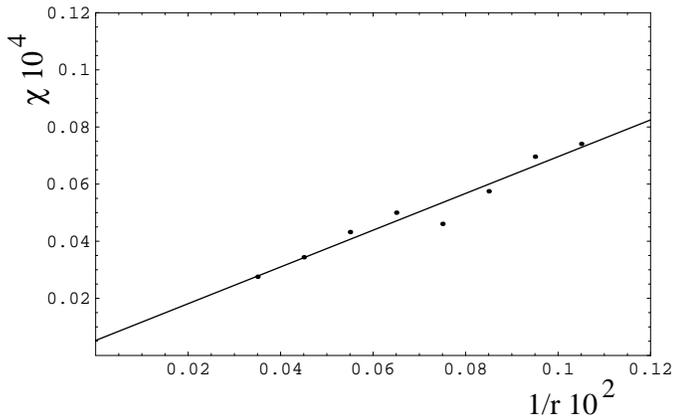,width=6cm,angle=270}
\caption{Susceptibility for stock price changes (slope of the price impact
function for small order imbalances) plotted against standard deviation 
$1/r$ of returns per trade. The linear relationship between the two 
quantities confirms the analogy to the fluctuation dissipation theorem 
$\sqrt{\langle (G_{\Delta t})^2\rangle } \sim  \chi$. The regression
coefficient of the linear fit is $R=0.975$.}
\label{flucdiss.f}
\end{figure}

Having classified all trades, we can give a sign to the volume of each
trade. A buyer initiated trade has a positive sign, a seller
initiated  trade a  negative
sign. As our analysis does not focus on the market microstructure, we
average both volume and price changes over a sampling time $\Delta t$.
This averaging is analogous to the concept of coarse graining in
physics, which also averages over microscopic model details and allows
the derivation of effective theories.  In the finance context, the
choice of the sampling time is influenced by the following
considerations.  If the sampling time is chosen too short, the effects
of individual trades will not be smoothed out, if it is chosen too
long it is not clear that there is still a causal relation between
order imbalance and price change: price changes will influence traders
and in this way have an effect on the order imbalance. For our
analysis we choose $\Delta t$ as five minutes. In this way we 
generate for each stock a time series of returns and order imbalances 
(measured relative to the total number of outstanding stocks of a company).

To visualize this information, we plot the average price change for a
given order imbalance against the order imbalance $\Delta Q$ (price
impact function) in Fig.\ref{chi.f}.  It agrees with the results of
\cite{private,response,KeKo}. Its characteristic form with a linear
part for small $\Delta Q$ and a less steep part for large $\Delta Q$
is described by a hyperbolic tangent \cite{private,response} and can
be well approximated by the functional form resulting from
Eq.(\ref{Langevin}).

Next, we sort the 5 minute intervals according to the average
fluctuation ($1/r$ in terms of the Langevin model) per trade and
calculate a price impact function for each fluctuation strength. We
calculate the slope in its linear region and thus generate a series of
susceptibilities $\chi$ depending on the local volatility. This data
set is shown in Fig.\ref{flucdiss.f}. The full line is a least square
fit to the data points. The good quality of the fit is illustrated by
the high regression coefficient $R=0.975$. Thus the empirical analysis
supports well the linear relationship between fluctuation strength and
susceptibility $\chi$  suggested by Eq.(\ref{flucdiss.eq}).

{\it Discussion:} The linear relation between fluctuation strength and
liquidity implies that large price changes are not only due to a large
volume of trades but also to a large price impact of a given volume.
In the soft spin model Eq.(\ref{Langevin}), the time changing
volatility is proportional to the square root of the number of trades
times the liquidity. In Ref.\cite{diffusion} it was shown that
changes in the trading frequency cannot account for the appearance of
large price fluctuations. Hence we conclude that large price
fluctuations are caused by changes in the market liquidity.

The inverse susceptibility appears as the $1/r$-term in our model
Eq.(\ref{Langevin}).  For this reason, large price fluctuations
associated with a large liquidity are described by the soft spin model
in a critical state.  In a critical spin model, even small changes in
the external magnetic field cause huge fluctuations in magnetization.
From this point of view, it is not large ``external'' influences which
cause large price fluctuations, but the strong response of the system
itself.

In summary, we have studied both theoretically and experimentally the
relation between the strength of stock price fluctuations and the
friction constant, which relates order imbalance and price changes.
We have modeled stock price dynamics on a coarse grained time scale by
a soft spin model and found an analogy to the fluctuation-dissipation
theorem in physics, which is supported by our empirical study.

Acknowledgements: We acknowledge several stimulating discussions with
V. Plerou, P. Gopikrishnan, X. Gabaix, and H.E. Stanley and thank them
for making available the results of \cite{response} prior to
publication.  We acknowledge helpful conversations with A. Kempf, T.
Nattermann, D. Stauffer, and L. Viceira.



\end{multicols}

\end{document}